\documentclass[12pt]{article}

\usepackage{hyperref}

\usepackage{amsmath} 
\usepackage{amssymb}
\usepackage{amsfonts}
\usepackage{amsthm}
\usepackage{graphicx}

\usepackage{amsmath} 
\usepackage{amssymb}
\usepackage{amsfonts}
\usepackage{amsthm}
\usepackage{bm} 
\usepackage{icomma} 

\usepackage{bbold}

\usepackage{mathtext} 				
\usepackage[T2A]{fontenc}			
\usepackage[utf8]{inputenc}

\usepackage{xcolor}

\theoremstyle{definition}

\begin{document}
	\begin{center}
		\Large
		\textbf{Long-time behavior of multi-level open systems  interacting with non-vacuum reservoirs}
		
		\large 
		\textbf{A.E. Teretenkov}\footnote{Department of Mathematical Methods for Quantum Technologies, Steklov Mathematical Institute of Russian Academy of Sciences,
			ul. Gubkina 8, Moscow 119991, Russia\\ E-mail:\href{mailto:taemsu@mail.ru}{taemsu@mail.ru}}
	\end{center}
	
	\footnotesize
		The model of multi-level open  quantum system  interacting with a non-vacuum reservoir in the rotating wave approximation is considered. We provide an exact integral representation for the reduced density matrix of the system.  For identical  uncorrelated reservoirs in diagonal states we have obtained the first perturbative correction for such dynamics in the Bogolubov–van Hove limit. We have shown that after initial state renormalization it can be completely described in terms of finite-dimensional semigroup. The method we provide can also be applied to the further orders of perturbation theory with  Bogolubov–van Hove scaling.
	\normalsize
	
\label{sec:intro}
\section*{Introduction}

The long-time behavior of open quantum systems is important. Only if the  timescale of the system dynamics separates from the reservoir timescale, then the open system becomes  a physical system having its own closed description of dynamics. Otherwise, it is just some degrees of freedom formally taken from the total system. The long-time behavior is closely related to the stochastic limit of the total unitary dynamics \cite{accardi2002quantum}, the possibility of  Markovian embedding of the density matrix dynamics  \cite{trushechkin2023long}, the long-time Markovianity \cite{teretenkov2021non, teretenkov2021long, teretenkov2023quantum} of the density matrix and multi-time correlation functions.

In this work, we consider the multi-level spin-boson-like model with rotating wave approximation (RWA) from \cite{teretenkov2019non}.  In \cite{teretenkov2021long} the long-time Markovian behaviour of this model was demonstrated, i.e. Markovian behavior of the density matrix and multi-time correlation functions at long times with additional renormalization due to initial non-Markovian behavior. It is well known \cite{accardi2002quantum} that this is the case for the model in the  Bogolubov–van Hove limit even without renormalization. But in this work we focus on the first non-trivial perturbative correction to this limit for non-vacuum and, moreover, non-factorizable initial state of the system and the reservoir.

We leave the derivation of RWA outside the scope of our work. But remark that RWA is also valid in the weak coupling and long-time regime \cite{fleming2010rotating, tang2013comparison, trubilko2020theory, burgarth2022one}. Thus, our approach does not break the domain of its validity.

\section*{Multi-level spin-boson-like model with RWA}

Recall the model from \cite{teretenkov2019non}. We consider the Hilbert space 	$\mathcal{H} \equiv (\mathbb{C}\oplus\mathbb{C}^{N}) \otimes \bigotimes_{j=1}^N \mathfrak{F}_b(\mathcal{L}^2(\mathbb{R})).$ Let $ | j \rangle , j = 0,  1, \ldots,  N $ be an orthonormal basis in $\mathbb{C}\oplus\mathbb{C}^{N} $,  $ | 0 \rangle $ is interpreted as a ground state.  $ \mathfrak{F}_b(\mathcal{L}^2(\mathbb{R})) $ are bosonic Fock spaces describing the reservoirs. Let $ | \Omega \rangle $ be a vacuum vector for the reservoirs. Let us also introduce the creation and annihilation operators which satisfy the canonical commutation relations: $ [b_{k,i}, b_{k',j}^{\dagger}] = \delta_{ij} \delta (k - k')$, $[b_{k,i}, b_{k',j}] = 0 $, $ b_{k,i} | \Omega \rangle = 0$.

We consider the system Hamiltonian of the  form  $ \hat{H}_S  =  0 \oplus H_S $, where $ H_S $  is an $ N \times N $ Hermitian matrix. The reservoir Hamiltonian is a sum of similar Hamiltonians of the free bosonic fields (with the same dispersion relation $ \omega_k $ and form factors $g_k$) and the interaction Hamiltonian has the RWA form
\begin{equation*}
	\hat{H}_B =\sum_{j=1}^N \int \omega(k) b_{k,j}^{\dagger} b_{k,j}  d k, \; \hat{H}_I = \sum_{j=1}^N \int \left(  g_{ k}^*  | 0 \rangle \langle j| \otimes b_{k,j}^{\dagger}+   g_{k}  | j \rangle \langle 0 | \otimes b_{k,j} \right) d k. 
\end{equation*}
The total dynamics of the system and the reservoirs is unitary
\begin{equation*}
	\rho(t) = e^{- i \hat{H} t} \rho(0) e^{ i \hat{H} t}
\end{equation*}
with the Hamiltonian $ \hat{H} = \hat{H}_S \otimes I + I \otimes \hat{H}_B + \hat{H}_I $. In contrast to \cite{teretenkov2019non, teretenkov2021long} we consider the initial condition of the form
\begin{equation}\label{eq:initCond}
	\!\!
	\rho(0) = \left(p \sigma \otimes|\Omega\rangle \langle \Omega| + (1-p) |0\rangle \langle 0| \otimes \int dk dk' \varrho_{ij}(k,k') b_{k,i}^{\dagger}|\Omega\rangle \langle \Omega| b_{k',j}\right),
\end{equation}
where $\sigma$ is a density matrix and
\begin{equation*}
	p \in [0,1], \quad \int dk \int dk' \varrho_{ij}(k,k') =1, \quad  \int dk \int dk' \varrho_{ij}(k,k') c_i^*(k) c_j(k') \geqslant 0.
\end{equation*}
Remark that  density matrix \eqref{eq:initCond} is non-factorized for $p \in (0,1)$, but it is still separable (non-entangled) for all $p \in [0,1]$. 

We will be interested in the reduced  dynamics of the multi-level system in the interaction picture
\begin{equation*}
	\rho_{SI}(t) \equiv \operatorname{Tr}_{B} \bigl(  e^{ i (\hat{H}_S \otimes I + I \otimes \hat{H}_B) t} \rho(t)  e^{- i (\hat{H}_S \otimes I + I \otimes \hat{H}_B) t} \bigr).
\end{equation*}

\section*{Integral representation of solution}

Since the RWA Hamiltonian preserves the total number of excitations, then the Schrödinger equation in the interaction picture has a solution of the form
\begin{equation}\label{eq:pureSolution}
	| \Psi(t)\rangle  = (\psi_0(0) \oplus | \psi(t)\rangle ) \otimes |\Omega\rangle + |0\rangle \otimes \int dk \psi_{k,j}(t)  b_{k,j}^{\dagger} |\Omega\rangle, \quad | \psi(t)\rangle \in \mathbb{C}^N.  
\end{equation}

By direct multi-level generalization of Lemma~1 in \cite{teretenkov2020integral} and transformation into the interaction picture, we obtain the following lemma.

\noindent\textbf{Lemma 1.}
\textit{
	Let the integrals 
	\begin{equation*}
		G(t)=\int d k\left|g_k\right|^2 e^{-i \omega(k) t}, \quad f_j(t)=\int d k g_k e^{-i \omega(k)  t} \psi_{k,j}(0), 
	\end{equation*}
	converge for all $t \in \mathbb{R}_{+}$ and define the continuous functions $G(t)$ and $f_j(t)$. And let us collect $f_j(t)$ in a single vector $|f(t) \rangle \in \mathbb{C}^N$ , $\langle j |f(t) \rangle = f_j(t)$, then $| \psi(t)\rangle$ is a  solution of the equation
	\begin{equation}\label{eq:integroDiff}
		\frac{d}{dt} |\psi(t)\rangle =  -i e^{i H_S t} |f(t)\rangle  - \int_{0}^{t} ds \; G(t-s) e^{i H_S (t-s)}  |\psi(s)\rangle.
	\end{equation}
}

Using Theorem 2.3.1 of \cite{burton2005volterra} we have the following lemma.

\noindent\textbf{Lemma 2.}
\textit{
	Let $G(t)$ be a function of exponential order and $V(t)$ be the solution of a Cauchy-like problem for the integro-differential equation
	\begin{equation*}
		\frac{d}{dt} V(t) = - \int_{0}^{t} ds \; G(t-s) e^{i H_S (t-s)} V(s), \qquad  V(0) = I,
	\end{equation*}
	then the solution of Eq.~\eqref{eq:integroDiff} has the form
	\begin{equation*}
		|\psi(t)\rangle =  V(t)|\psi(0)\rangle - i \int_0^t ds   V(t-s) e^{i H_S s}  |f(s)\rangle .
	\end{equation*}
}

The reduced density matrix corresponding to the pure state defined by Eq.~\eqref{eq:pureSolution} has the block form
\begin{equation*}
	\rho_{S I}(t)=
	\begin{pmatrix}
		| \psi_0(0)|^2 + \|\psi(0)\|^2 - \|\psi(t)\|^2 & \psi_0(0)\langle\psi(t)| \\
		\psi_0^*(0)|\psi(t)\rangle & |\psi(t)\rangle\langle\psi(t)|
	\end{pmatrix}.
\end{equation*}
Then decomposing an arbitrary initial state \eqref{eq:initCond} into a sum of projectors to the pure states \eqref{eq:pureSolution} we obtain the following theorem.\\[-0.45cm]

\noindent\textbf{Theorem 1.}
\textit{
	Assuming that the conditions of lemmas 1 and 2 are satisfied, let us define
	\begin{equation*}
		K(s,s') \equiv  \int dk dk' \varrho_{jj'}(k,k')  e^{-i( \omega(k) s -  \omega(k') s')} g_k  g_{k'}^* |j\rangle \langle j' |,
	\end{equation*}
	then
	\begin{equation}
		\rho_{S I}(t)=
		\begin{pmatrix}
			(\rho_{SI}(0))_{00} + \operatorname{Tr}( (\rho_{SI}(0))_{ee}- (\rho_{SI}(t))_{ee}) &(\rho_{SI}(0))_{0e} 	V^{\dagger}(t)\\
			V(t)(\rho_{SI}(0))_{e0} & (\rho_{SI}(t))_{ee}
		\end{pmatrix},
		\label{eq:rhoSI}
	\end{equation}
	where
	\begin{align}
		(& \rho_{SI}(t))_{ee} =p  V(t)  \sigma_{ee} V^{\dagger}(t) \nonumber\\
		&- (1-p) \int_0^t ds \int_0^t ds'  V(t-s) e^{i H_S s}   K(s,s')  e^{-i H_S s'} V^{\dagger}(t-s'). \label{eq:exitedIntegral}
	\end{align}
}

\section*{Long-time behavior for identical uncorrelated reservoirs in diagonal states}

Now let us assume that $	\varrho_{jj'}(k,k')$ has a particular form
\begin{equation*}
	\varrho_{jj'}(k,k') =  \varrho(k) \delta_{jj'} \delta(k-k'),
\end{equation*}
i.e. the reservoirs are uncorrelated and have identical initial states, which are  diagonal in the momentum basis. Let us define the non-vacuum spectral density 
\begin{equation*}
	J_{\varrho}(\omega) \equiv \int dk \delta(\omega(k) - \omega) \varrho(k) |g_k|^2,
\end{equation*}
then $K(s,s')$ becomes proportional to the identity matrix $I$:
\begin{equation}\label{eq:uncorrelatedKernerl}
	K(s,s') =  \int d\omega J_{\varrho}(\omega) e^{-i\omega ( s -  s') } I.
\end{equation}

Now let as consider the perturbation theory with Bogolubov--van Hove scaling $g_k \rightarrow \lambda g_k$ (weak coupling) and $t \rightarrow \lambda^{-2} t$ (long times) with $\lambda \rightarrow +0$. Remark that $g_k \rightarrow \lambda g_k$ leads to scaling $ G(t) \rightarrow  \lambda^2 G(t)$ and $	K(s,s') \rightarrow  \lambda^2 	K(s,s')  $. If we  highlight the explicit dependence on $\lambda$ form $\lambda g_k $ in  Eq.~\eqref{eq:rhoSI} as  $	\rho_{S I}(t ;\lambda)$, then we are interested  the perturbative expansion of $\rho_{S I}(\lambda^{-2} t ;\lambda)$, the first terms of which are given by the next theorem.\\[-0.4cm]

\noindent\textbf{Theorem 2.}
\textit{
	Let  the conditions of Theorem 1 satisfied and let $K(s,s')$ be defined by \eqref{eq:uncorrelatedKernerl} with $J_{\varrho}(\omega)$ from the Schwartz space $\mathcal{S}(\mathbb{R})$. Let  $\tilde{G}(p)$ be the Laplace transform of $G(t)$ and let us assume that $\tilde{G}(- i H_S) >0 $. Then for fixed $t$ and $\lambda \rightarrow +0$ we have
	\begin{equation}\!\!\!\!
		\rho_{S I}(\lambda^{-2} t ;\lambda) =
		\begin{pmatrix}
			(1 - \operatorname{Tr}((\rho_{SI}(\lambda^{-2} t ;\lambda))_{ee}) &(\rho_{SI}(0))_{0e}  r^{\dagger}(\lambda)  e^{L^{\dagger}(\lambda)  t}\\
			e^{L(\lambda)  t} r(\lambda)(\rho_{SI}(0))_{e0} & 	 (\rho_{SI}(\lambda^{-2} t ;\lambda))_{ee}
		\end{pmatrix} + o(\lambda^{2}),\label{eq:rhoSIpert}
	\end{equation}
	where $L(\lambda) = - \tilde{G}(- iH_S) + \lambda^2 \tilde{G}'(- iH_S) \tilde{G}(- iH_S)$, $r=1-\lambda^2 \tilde{G}^{\prime}\left(-i H_S\right)$,
	\begin{equation*}
		(\rho_{SI}(\lambda^{-2} t ;\lambda))_{ee} = e^{L(\lambda)  t} ( r(\lambda) \bigl(\rho_{SI}(0))_{ee} r^{\dagger}(\lambda) -  (\rho_{SI}(+\infty))_{ee}\bigr)   e^{L^{\dagger}(\lambda)  t} 
		+  (\rho_{SI}(+\infty))_{ee},
	\end{equation*}
	\begin{equation*}
		(\rho_{SI}(+\infty))_{ee} =\frac{ 2 \pi}{L(\lambda) + L^{\dagger}(\lambda)}   r(\lambda)  \left(J_{\varrho}(H_S)  + i \lambda^2 J_{\varrho}'(H_S) \frac{L^{\dagger}(\lambda)-L(\lambda)}{2} \right)   r^{\dagger}(\lambda) .
	\end{equation*}
}
\\[-0.3cm]
\noindent\textit{Proof.} Using Theorem 2 from \cite{teretenkov2021long} we have $ V(\lambda^{-2}( t-s); \lambda) =  e^{L(\lambda)  t} r(\lambda) + o(\lambda^2)$.
By Eq. \eqref{eq:uncorrelatedKernerl} and  \cite[Theorem 3.1.]{pechen2002quantum} we have
\begin{align*}
	e^{i H_S  \lambda^{-2}s}  \frac{1}{\lambda^2} K(\lambda^{-2} s, \lambda^{-2} s')  e^{-i H_S \lambda^{-2}s'} =&2 \pi( J_{\varrho}(H_S) \delta(s-s') \\
	&- i \lambda^2 J_{\varrho}'(H_S) \delta'(s-s') ) +o(\lambda^2),
\end{align*}
then the integral in \eqref{eq:exitedIntegral} after  Bogolubov–van Hove scaling takes the form
\begin{align*}
	&\lambda^2 \int_0^{\lambda^{-2} t} ds \int_0^{\lambda^{-2} t} ds'  V({\lambda^{-2} t}-s;\lambda) e^{i H_S s}   K(s,s')  e^{-i H_S s'} V^{\dagger}({\lambda^{-2} t}-s';\lambda)\\
	&= \int_0^{ t} ds \int_0^{ t} ds'  V\biggl( \frac{ t-s}{\lambda^2};\lambda\biggr) e^{i  \frac{s}{\lambda^2} H_S }  \frac{1}{\lambda^2} K(\lambda^{-2} s, \lambda^{-2} s')  e^{-i  \frac{s'}{\lambda^2} H_S}  V^{\dagger}\biggl( \frac{ t-s'}{\lambda^2};\lambda\biggr)\\
	&= 2 \pi  \frac{e^{(L(\lambda) + L^{\dagger}(\lambda))t} -1}{L(\lambda) + L^{\dagger}(\lambda)}  r(\lambda)  \left(J_{\varrho}(H_S)  + i \lambda^2 J_{\varrho}'(H_S) \frac{L^{\dagger}(\lambda)-L(\lambda)}{2} \right)   r^{\dagger}(\lambda) +  o(\lambda^2).
\end{align*}
Substituting it into Eqs.~\eqref{eq:rhoSI}-\eqref{eq:exitedIntegral} we obtain Eq.~\eqref{eq:rhoSIpert}.

\section*{Conclusions}

We have considered the long-time behavior of the model in the simplest non-trivial case of similar uncorrelated reservoirs in diagonal states. But we remark that Theorem 1 is more general and allows one to consider correlated reservoirs in different states as well. Moreover, following \cite{lonigro2022quantum} it is possible to generalize our results to the model with different dispersion relations and form-factors, but using matrix-valued function $G(t)$. Long-time behavior of such models is an interesting direction for the further study, since it can describe flows and currents through the system between different reservoirs. 

In Theorem 2 we have considered only the first non-trivial correction to the  Bogolubov–van Hove limit. However, analogous to    \cite{teretenkov2021non}, under additional but still very general conditions, one can obtain similar semigroup approximation of $V(\lambda^{-2}( t-s); \lambda)$ in all the orders of perturbation theory. The multipole expansion of the integral kernel can  also be done in any order following \cite{pechen2002quantum}. Thus, Theorem 2 can  also be generalized to all the orders of perturbation theory. 

Due to the fact that Theorem 2  generalizes results of \cite{teretenkov2021non, teretenkov2021long} to non-factorized initial states describing the perturbative dynamics of the reduced density matrix in terms of semigroup with renormailziation of initial conditions, then it can be considered as long-time Markovian behavior. But it is still an open question: are there some renormalized analogues of the regression formula for multi-time correlation functions in our case? The results of \cite{teretenkov2021non, teretenkov2021long, lonigro2022quantum} for the vacuum reservoirs suggest that there should be, if the reduced density matrix dynamics is described by a semigroup.

Another direction of the further study is to consider the regimes where long-time dynamics is not described by semigroups \cite{teretenkov2021non, trushechkin2023long, basharov2021cooperative}. 

\label{sec:acknowledgement}
\section*{Acknowledgement}
The author is grateful to the Referee for the important remarks.

\bibliographystyle{unsrt}
	\bibliography{longtime}
\end{document}